\documentclass{article}
\usepackage{amsmath}
        \newcounter{eqnletter}[equation]
\catcode`\@=11 \@addtoreset{equation}{section} \catcode`\@=12

\usepackage[dvips]{graphicx}
\usepackage{color}
\usepackage{epsfig}

\begin{document}
\centerline  {\LARGE A random matrix model for the density of states of } 
\centerline {\LARGE jammed soft spheres with applied stress}
\vskip .8cm
{\centerline {Mario Pernici$^{*}$
\footnote{mario.pernici@mi.infn.it}}}
\vskip .3 cm
        {\centerline  {$^{*}$ Istituto Nazionale di Fisica Nucleare,
        Sezione di Milano,}}
{       \centerline  {  Via Celoria 16, 20133 Milano, Italy}}

\vskip 1cm {\centerline{\textbf{Abstract}} }
                \vskip .4 cm

We investigate the addition of applied stress to a random block matrix model 
introduced by Parisi to study the Hessian matrix of soft spheres near the jamming point.
In the infinite dimensional limit the applied stress translates
the spectral distribution to the left, leading to a stability constraint.
With negative stress, as in the case of a random network of stretched elastic springs, 
the spectral distribution is translated to the right, and the density of states 
has a peak before the plateau.

\vskip 0.3cm
Physics Subject Headings:

Physical systems: Amorphous materials, Disordered systems, Random graphs

Techniques: Random matrix theory

\section{Introduction}

The eigenvalue spectrum of sparse random matrices has many applications in 
physics \cite{rev1}.

Amorphous solids close to mechanical instability have an excess of low frequency 
modes with respect to the Debye law, often referred to as the boson peak \cite{phil}.
A similar phenomenon is present in granular matter near the jamming point;
in \cite{ohern}  a system of elastic spheres with random close
packing is considered, in which the network of contacts between the spheres forms a 
random graph with average connectivity $Z$. Using numerical simulations with a system of
particles
it is found in \cite{ohern} that at the isostatic jamming point $Z = 2 d$ \cite{maxw} 
the density of states (DOS) has a plateau around zero frequency. Increasing the
connectivity the plateau starts at higher frequency.
An explanation for the presence of the plateau at the isostatic point is given in
\cite{wya}.

As observed in \cite{alex}, 
positive stress (compression) leads to lower frequencies of the soft modes,
decreasing the stability, so the system jams at $Z > 2 d$;
negative stress (stretching) leads to higher frequencies, increasing the stability,
so the system jams at $Z < 2 d$, as in gels.
In the framework of nonaffine lattice dynamics \cite{zs} the modification of the Maxwell 
rigidity criterium is studied in \cite{crz}.

In \cite{degiuli} they model amorphous solids with a random elastic network 
with average connectivity $Z$, 
obtained by random dilution of a regular lattice in three
dimensions; they compute the Green function and the DOS
for the dynamical matrix in presence of applied stress.
They find that there is a stability region in the $Z, e$ plane, where $e$ is the
applied compressive strain; they discuss both repulsive particles ($e > 0$) and amorphous
solids with negative strain coefficient, in particular to describe silica at low
frequencies.

In \cite{ml} it has been suggested to study the boson peak in amorphous solids
using a class of Laplacian random matrices, called diagonal-dominant random matrices.
In \cite{pari} a Laplacian sparse random block matrix on random regular graphs has been
considered, as a mean field approximation of the Hessian matrix of a system of elastic 
spheres with random close packing; it is assumed that the network of contacts is a 
random regular graph with degree $Z$, and that the contact interactions are
randomly oriented;
in this way the soft modes are due only to disorder, not
to the Goldstone modes of an underlying lattice.
The applied stress is neglected, so this model describes
a random network of unstretched elastic springs.
Using the cavity method it has been shown in \cite{pari}
that, in the limit $d \to \infty$ with $t = \frac{Z}{d}$ fixed,
the spectral distribution is the Marchenko-Pastur distribution \cite{marcpas}
with parameter $t$; the DOS in low dimensions and on a class of almost
regular graphs has been studied with the cavity method and simulations in \cite{benet}.

In \cite{cic1} it has been conjectured that the same limiting distribution
is obtained in the case of random Erd\"os-Renyi graphs;
in \cite{pc} this has been proven observing that the only contributions to the moments
in this limit are due to walks which have a noncrossing sequences \cite{ncr} of steps.
Another derivation of the spectral distribution has been obtained in \cite{dkl}.
In \cite{pern2} it is observed that the proof in \cite{pc} holds also
in the case in which random regular graphs instead of 
Erd\"os-Renyi graphs are used.

In this letter we extend the mean field model introduced in \cite{pari}, adding 
applied stress. 
We prove that in the limit $d \to \infty$ with $t=\frac{Z}{d}$ constant
and applied stress times $d$ constant, 
the spectral distribution is a translated Marchenko-Pastur distribution;
from this  it follows that there is a stability region for the jammed phase.
Simulations with random matrices in low dimensions show that in this region
there is a small fraction of negative eigenvalues,
decreasing as $d$ increases.
Changing the sign of the applied stress the Marchenko-Pastur distribution is
translated in the opposite direction. In the hypostatic region
$t < 2$ there is a delta-function peak
in the DOS; for $t - 2$ small and positive, there is a peak in presence of enough
applied stress. Simulations show that for $t < 2$ the delta-function peak 
becomes in low dimensions a peak with width depending on the applied stress.

In Section II we present the random block matrix model for an amorphous solid with
applied stress; in subsection II.1 we study the case of soft spheres,
in subsection II.2 the case of negative applied stress.

\section{A random block matrix model for amorphous solids near the isostatic point}

Following Alexander \cite{alex} (see also \cite{wyart2} for a review), consider 
a system of particles
with a two-particle potential $V(r)$; let $\vec{R}_i$ be the position of a particle,
 $\delta \vec{R}_i$ its displacement,
$r_{i,j} = ||\vec{R}_i - \vec{R}_j||$ and 
$\vec{n}_{i,j} = \frac{\vec{R}_i - \vec{R}_j}{r_{i,j}}$.
The change in the energy with
respect to the equilibrium position is, to second order in the displacements,
\begin{eqnarray}
	\delta E &=& \frac{1}{2} \sum_{i,j} c_{i,j}
[(\delta \vec{R}_j - \delta \vec{R}_i)^\perp]^2 +
k_{i,j} [(\delta \vec{R}_j - \delta \vec{R}_i)\cdot \vec{n}_{i,j}]^2
\label{de1} \\
	&&c_{i,j} = \frac{V'(r_{i,j}^{\textrm{eq}})}{r_{i,j}^{\textrm{eq}}}; \quad
	k_{i,j} = V''(r_{i,j}^{\textrm{eq}})
\label{de2}
\end{eqnarray}
where $(\delta \vec{R}_j - \delta \vec{R}_i)^\perp$ indicates the projection
of $\delta \vec{R}_j - \delta \vec{R}_i$ 
on the (hyper)plane orthogonal to $\vec{n}_{i,j}$.
The first term in Eq. (\ref{de1}) is called initial stress or applied stress.
Let us choose the scale such that the stiffness $k_{i,j}$ of particles in contact 
is $1$ in average.

Taking the approximation
\begin{eqnarray}
c_{i,j} &&= \frac{c}{d} \, \alpha_{i,j} \nonumber \\
k_{i,j} &&= \alpha_{i,j}
\label{pram}
\end{eqnarray}
where $\alpha_{i,j} = 1$ if the particles are in contact and zero otherwise,
the off-diagonal block of the Hessian matrix 
$\frac{\partial^2 U}{\partial R_i^\alpha \partial R_j^\beta}$, with 
$\alpha, \beta = 1, \cdots,d$, is $-\alpha_{i,j} X_{i,j}$, where
\begin{equation}
X_{i,j} = \big(1 - \frac{c}{d}\big) |n_{i,j}><n_{i,j}| + \frac{c}{d} I_d
\label{xr2}
\end{equation}
and $I_d$ is the identity matrix in $d$ dimensions.

Following \cite{pari}, consider a mean field approximation,
in which  $n_{i,j}$ are random independent uniformly distributed vectors
on the unit sphere in $d$ dimensions and the graph of contacts of the spheres
forms a random regular graph of degree $Z$.

Consider a Laplacian random block matrix whose off-diagonal blocks
have the form $L_{i,j} = -\alpha_{i,j} X_{i,j}$, with $i,j=1,\cdots, N$;
the diagonal blocks have the form
$L_{i,i} = \sum_{j \neq i} \alpha_{i,j} X_{i,j}$,
where $\{\alpha_{i,j} = \alpha_{j,i}\}$, with $i \neq j$, form a set of independent
identically distributed random variables with the probability density
\begin{eqnarray} 
P(\alpha)= \frac{Z}{N} \delta (\alpha
-1)+\left(1-\frac{Z}{N}\right) \delta(\alpha).\qquad \qquad \label{d.2}
\end{eqnarray}

The generating function of the moments is
\begin{equation}
        f(x) = \lim_{N \to \infty} \frac{1}{Nd} \sum_{n \ge 0} x^n
        \sum_{j=1}^N <\textrm{tr} (L^n)_{j,j}>
\end{equation}
where $\textrm{tr}$ is the $d$-dimensional trace.
The only contributions to the moments for $N \to \infty$ come from walks on trees.

We want to compute the spectral distribution $\rho^{(c,t)}$ of $L$ in the limit 
$d \to \infty$ with $t = \frac{Z}{d}$ and $c$ fixed.

In \cite{pc} it has been proven that for $c = 0$ the spectral distribution
is the Marchenko-Pastur distribution $\rho_{\textrm{MP}}^{(t)}$ with parameter $t$.
In \cite{pern2} it has been shown that this proof extends to the case of a random
regular graphs, as found previously \cite{pari}.

Consider the contribution of a walk on a tree to a moment of $\rho^{(c,t)}$, 
and an edge $(i,j)$ in it. 
The applied stress term $-\frac{c}{d}\alpha_{i,j} I_d$ in $L_{i,j}$, 
with $i \neq j$, gives a vanishing contribution in the infinite dimensional limit: 
in fact the walk must pass at least another time
through the edge $(i,j)$, so the above term can be replaced
by $-\frac{c}{d} I_d$.
The remaining average is finite in the case of a walk with noncrossing sequence of
steps, giving a factor $t$ for each edge in the walk, vanishing otherwise in the
limit $d \to \infty$;
therefore the average vanishes in this limit.

The term in $L_{i,i}$ containing $\alpha_{i,j}$ is
$\alpha_{i,j}\big((1 - \frac{c}{d}) |n_{i,j}> <n_{i,j}| + \frac{c}{d} I_d \big)$.
If the walk does not pass
again through the edge $(i,j)$, one can average 
$\alpha_{i,j}$ in $\alpha_{i,j} \frac{c}{d} I_d$
obtaining $\frac{c t}{N} I_d$ and 
$\sum_j \alpha_{i,j} \frac{c}{d} I_d = c t I_d$ in the limit;
otherwise it gives vanishing contribution in the infinite dimensional limit
as in the case of $L_{i,j}$ considered above.
A contribution to the moments, with a term $\alpha_{i,j} |n_{i,j}> <n_{i,j}|$
coming from $L_{i,i}$, gives a finite or zero contribution in the limit, so
insertions of $\alpha_{i,j} \frac{c}{d} |n_{i,j}> <n_{i,j}|$ give zero contribution to 
the moments in this limit.
Therefore the moments of $L$ are equal to the moments of $L(c=0) + c t I_{Nd}$
in this limit; hence the spectral distribution is
\begin{equation}
\rho^{(c,t)}(\lambda) = \rho^{(t)}_{\textrm{MP}}(\lambda - c t)
\label{densM}
\end{equation}
where $\rho^{(t)}_{\textrm{MP}}$ is the Marchenko-Pastur distribution.
Since the domain of $\rho_{MP}^{(t)}$ is $[a_-(t), a_+(t)]$, where 
\begin{equation}
	a_\pm(t) = (\sqrt{t} \pm \sqrt{2})^2 
\label{apm}
\end{equation}
plus the point $\lambda = 0$ in the case $t < 2$,
the domain of $\rho^{(c,t)}$ is $[a_-(t) + c t, a_+(t) + c t]$, 
plus the point  $\lambda = c t$ in  the case $t < 2$.

The same spectral distribution holds using random regular graphs of degree $Z$ 
instead of Erd\"os-Renyi graphs \cite{pern2}.

\subsection{Soft spheres with applied stress}
In the case of soft spheres with two-particle potential
$V(r_{i,j}) = \frac{1}{2}(\sigma - r_{i,j})^2 \alpha_{i,j}$,
where $\sigma$ is the diameter of the spheres and
$\alpha_{i,j} = \theta(\sigma - r_{i,j})$ is $1$ 
if the particle are in contact, zero otherwise, one has 
$c_{i,j} = \frac{r_{i,j}^{\textrm{eq}} - \sigma}{r_{i,j}^{\textrm{eq}}}\alpha_{i,j}$
and $k_{i,j} = \alpha_{i,j}$.
So $c_{i,j} \le 0$, the applied stress is a compression term; let us set in 
Eq. (\ref{pram})
\begin{equation}
c = - R^2
\label{cm}
\end{equation}

Stability requires that $\rho^{(-R^2, t)}$ has positive domain; 
since for $t < 2$ there is a delta function
term in $\lambda = c t$, it must be $t \ge 2$ if $R \neq 0$.
The spectral distribution becomes non-vanishing in $\lambda = a_-(t) + c t$, so that
the stability condition is
\begin{equation}
	R \le R_0(t) \equiv \frac{\sqrt{t} - \sqrt{2}}{\sqrt{t}}\,;\quad
	t \ge \frac{2}{(1 - R)^2}
\label{stab}
\end{equation}
In particular stability requires $R=0$ for $t=2$.
We will restrict to $t > 2$ in the rest of the section.

In the notation in \cite{degiuli}, in $d=3$, the compressive strain factor is 
$e = -\frac{c}{d} = \frac{R^2}{d}$; $Z_c = 2 d$.
The  compressive strain factor at the boundary of the stability region is
\begin{equation}
	e_c = -\frac{R_0^2(t)}{d} = \frac{(\sqrt{Z} - \sqrt{Z_c})^2}{d Z}
\end{equation}
For $\delta Z = Z - Z_c$ small,
\begin{equation}
	e_c = \frac{1}{4 d Z_c^2} (\delta Z)^2 + O((\delta Z)^3)
\end{equation}
in agreement with $ e_c \sim (\delta Z)^2$ in \cite{degiuli}.

The DOS $D^{(c,t)}(\omega) = 2 \omega \rho^{(c,t)}(\omega^2)$, 
in the limit $d \to \infty$ with $t = \frac{Z}{d}$ fixed, is
\begin{eqnarray}
	D^{(c,t)}(\omega) &=& \frac{\omega}{2\pi (\omega^2 - c t)} 
	\sqrt{(a_{+}(t) + c t - \omega^2)(\omega^2 - c t - a_{-}(t))} \nonumber \\
	&&\sqrt{a_{-}(t) + c t} \le \omega \le \sqrt{a_{+}(t) + c t}
\label{dmr}
\end{eqnarray}
which for $c=0$ is the Marchenko-Pastur DOS with parameter $t$.

For $R < R_0(t)$, $D^{(c,t)}$ has a gap of size $\sqrt{(R_0^2(t) - R^2) t}$.

In finite dimensions the spectral density of a Laplacian random block matrix 
on Erd\"os-Renyi 
graphs is not the same as the one for random regular graphs; in low dimensions
the former has an excess of soft modes due to rattlers, so Erd\"os-Renyi graphs
are not suitable to investigate the spectral distribution of the Laplacian block matrix
near the jamming point.
On the other hand, also regular graphs are not suitable, since e.g. in $d=3$
one cannot get closer to the isostatic point than $t=\frac{7}{3}$.
Following \cite{benet}, we will use in simulations in the hyperstatic region the sum
of a random regular graph with degree $2 d$ and an Erd\"os-Renyi graph with degree
$(t-2)d$ to obtain a random graph with average degree $Z$.

\begin{figure*}[hbt!]
\begin{center}
\epsfig{file=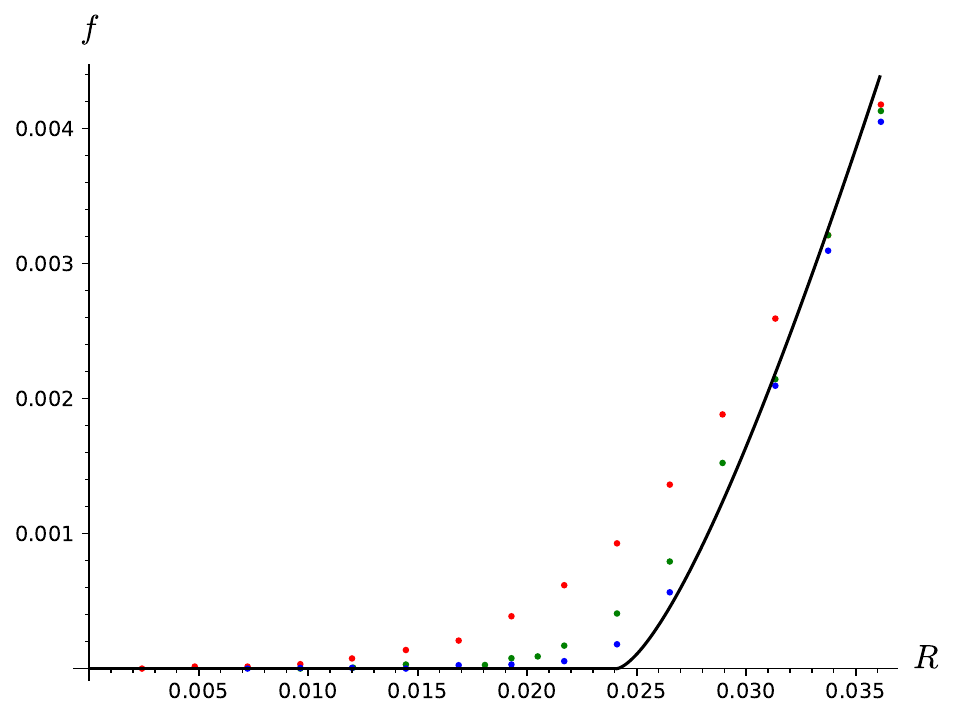,  width=7.00cm}
\caption{Plots of relative number $f$ of negative energy states 
	against $R$, for $t = 2.1$.
The simulations are done on random graphs with $1000$ nodes in 
	$d=2$ (red), $d=3$ (green) and $d=4$ (blue), respectively
with $200$, $100$ and $50$ random graphs.
The black line is $f$ in the infinite dimensional case.
}
\label{fneg}
\end{center}
\end{figure*}

In finite dimensions the system is unstable under applied stress: there
is a small number of negative eigenvalues in the spectral distribution.
This is due to the fact that for $R = 0$ there is a quasi-gap
\cite{benet}, so that applying pressure the few modes around $\lambda=0$ are
shifted to negative values.
We made simulations with random matrices\footnote{The matrix eigenvalues are computed 
with Sage \cite{sage} 
using  numpy's linalg.eigvalsh . }
in the case $t=2.1$;
in Fig. \ref{fneg} we plot the relative number of negative energy states
against $R$ in $d=2, 3$ and $4$. As $d$ increases, it decreases towards zero
for $R \le R_0$.

\begin{figure*}[hbt!]
\begin{center}
\epsfig{file=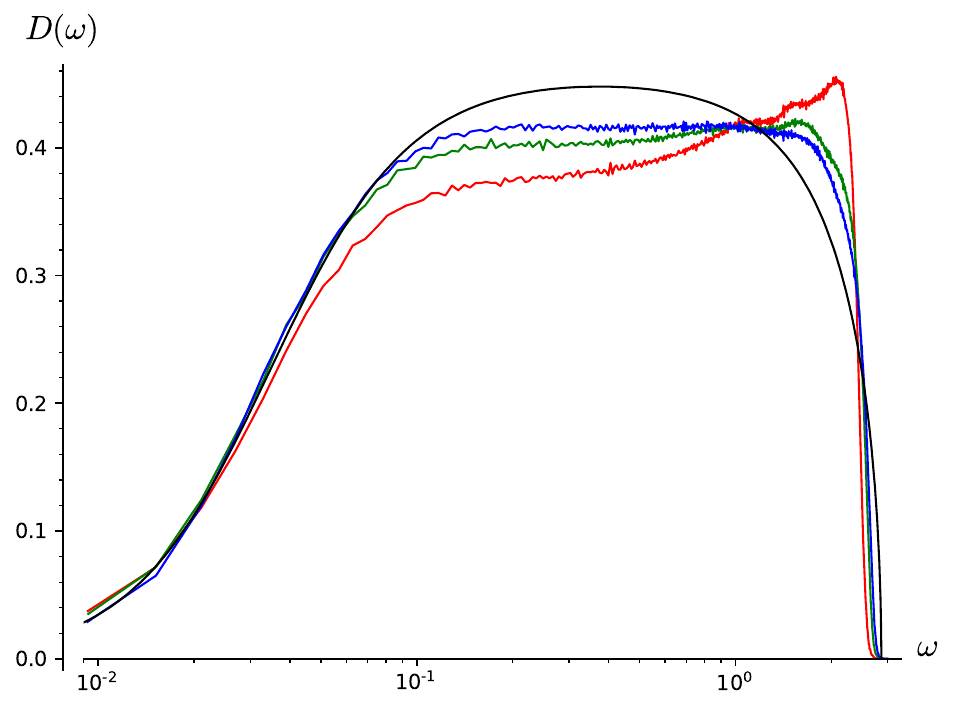,  width=5.00cm}
\epsfig{file=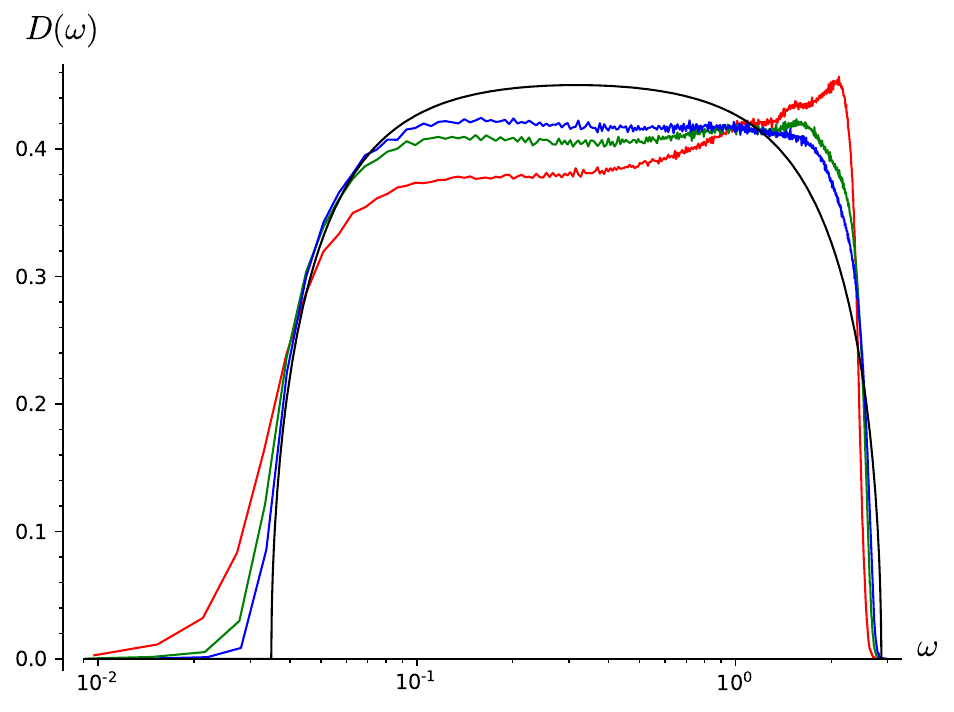,  width=5.00cm}
	\caption{Plots of simulations of the DOS for $t=2.1$, $R=0.0241$ (left) and 
	$R=0$ (right)
	in $d=2$ (red), $3$ (green) and $4$ (blue) respectively with
	$2000$, $1000$ and  $500$ random graphs with $1000$ nodes.
	The black line is the infinite dimensional solution.
}
\label{figt21r}
\end{center}
\end{figure*}

In Fig. \ref{figt21r} we plot $D^{(-R^2,2.1)}_d$ obtained with simulations in
$d=2, 3$ and $4$,
neglecting the negative eigenvalues present for $R \neq 0$; 
in the left hand figure $R = 0.0241 \approx R_0(2.1)$, in the right hand figure $R = 0$.
In both figures the $D^{(-R^2,t)}_d$ approaches $D^{(-R^2,t)}$ as $d$ increases.

\newpage
\subsection{Negative applied stress}
In the case of a random network of stretched elastic springs,
in the approximation Eq. (\ref{pram}) we can set
\begin{equation}
c = R^2
\label{cp}
\end{equation}
From Eq. (\ref{densM}) it follows that the spectrum is always positive,
so that there is no stability condition.
As observed in \cite{alex, crz, wyart2}, when the applied stress is negative,
one can have stable solids with connectivity less
than $2 d$. In this model, one can have therefore stability with $t < 2$.
The continuous part of the DOS in the infinite dimension limit is Eq. (\ref{dmr});
its singular part is
$\frac{2-t}{2} \theta(2 - t) \delta(\omega - R\sqrt{t})$, 
where $\theta(x)$ is the Heaviside step function.

At $t=2$ and $R > 0$, $D^{(R^2, 2)}$ has an integrable singularity
in $\omega_1 = \sqrt{2} R$, for $\omega$ close to $\omega_1$ one has 
$\int_{\omega_1}^\omega d\omega D^{(R^2, 2)}(\omega)\approx \frac{\sqrt{2}}{\pi} \sqrt{\omega^2 - 2 R^2}$.

The peak for $t > 2$, with $t-2$ small,
starts at $R$ a bit larger than $R_0(t)$ defined in Eq. (\ref{stab}).
The position of the peak can be computed approximately for $R$ small
finding the stationary points of the approximate DOS obtained
replacing $a_+(t) + c t - \omega^2$ with $a_+(t)$ in Eq. (\ref{dmr}).
For $R > \sqrt{1 + \sqrt{2}}\, R_0(t)$,
the position $\omega_p$ and the height of the peak for $t > 2$ are given by
\begin{eqnarray}
	\omega_p \simeq &&
	\sqrt{t}\, R \, \sqrt{ \frac{R^2 + R_0^2(t)}{R^2 - R_0^2(t)}  } \nonumber \\
	D^{(R^2,t)}(\omega_p) \simeq &&
	\frac{(R^2 + R_0^2(t))(\sqrt{t} + \sqrt{2})}{4\pi R R_0(t)}
	\label{omp}
\end{eqnarray}

\begin{figure*}[hbt!]
\begin{center}
\epsfig{file=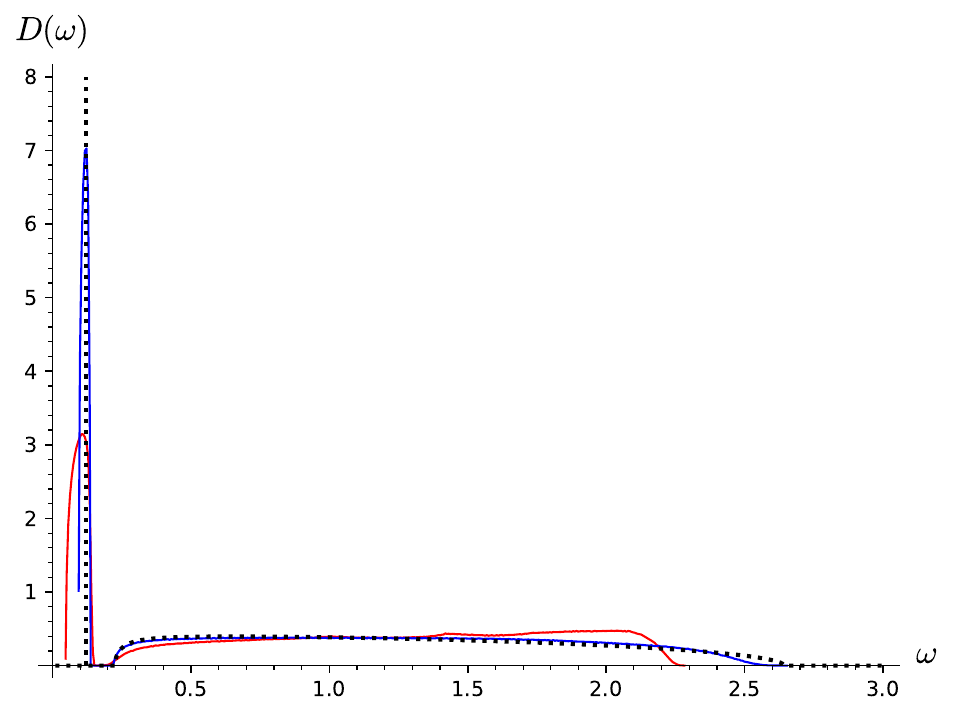,  width=5.00cm}
\epsfig{file=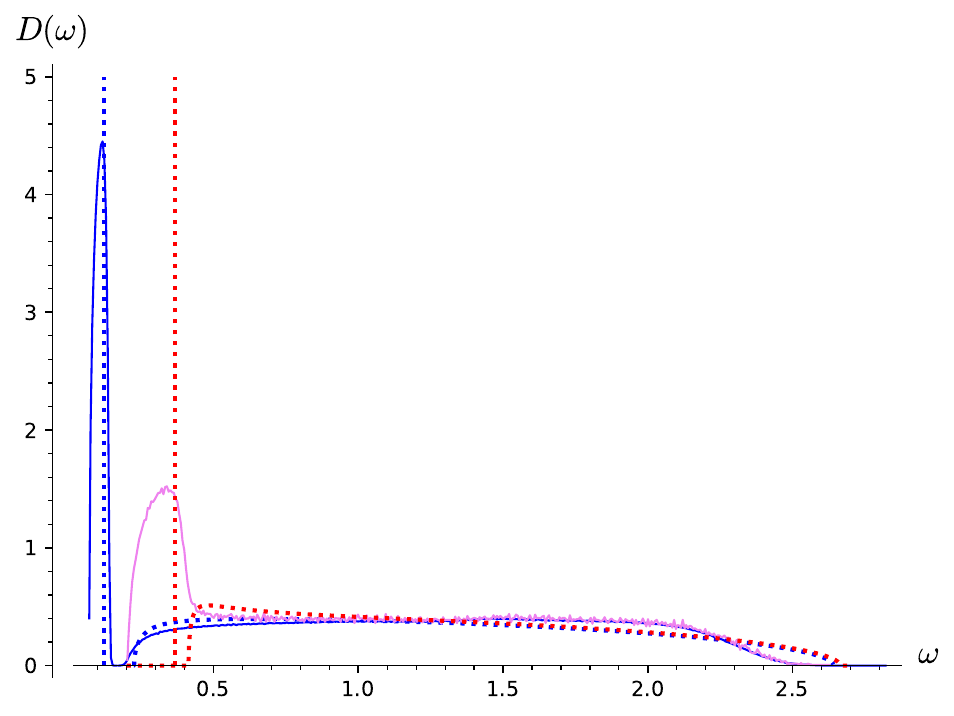,  width=5.00cm}
\caption{ Left: DOS with $t=1.5$ and $R=0.1$ obtained in simulations in 
	$d=2$ (red) and $6$ (blue)
with random graphs with $1000$ nodes, respectively with ensembles of $1000$ random
regular graphs with degree $3$ and $100$ random graphs with degree $9$.
Right: Plots of simulations of the DOS in $d=3$ for $t=1.5$, $R=0.1$ (blue), 
$0.3$ (red) with $500$ random graphs with $1000$ nodes.
The dashed lines represent the infinite dimensional solution.
        }
\label{figt15d}
\end{center}
\end{figure*}

In finite dimensions the delta-function present in $d=\infty$ for $t < 2$ is smeared;
increasing the dimension
the peak approaches the delta-function term. For instance we plot in 
the left hand figure in Fig. \ref{figt15d}
the DOS with $t=1.5$ and $R=0.1$ obtained in simulations in $d=2$ and $6$.

Increasing $R$ the peak corresponding to the delta function approaches and finally 
merges with the band corresponding to the continuous part of the DOS in infinite
dimensions,
see the right hand figure in Fig. \ref{figt15d};
in the simulations for this case we used the sum of random regular graphs with
degree $4$ and Erd\"os-Renyi graphs with average degree $0.5$.

\begin{figure*}[hbt!]
\begin{center}
\epsfig{file=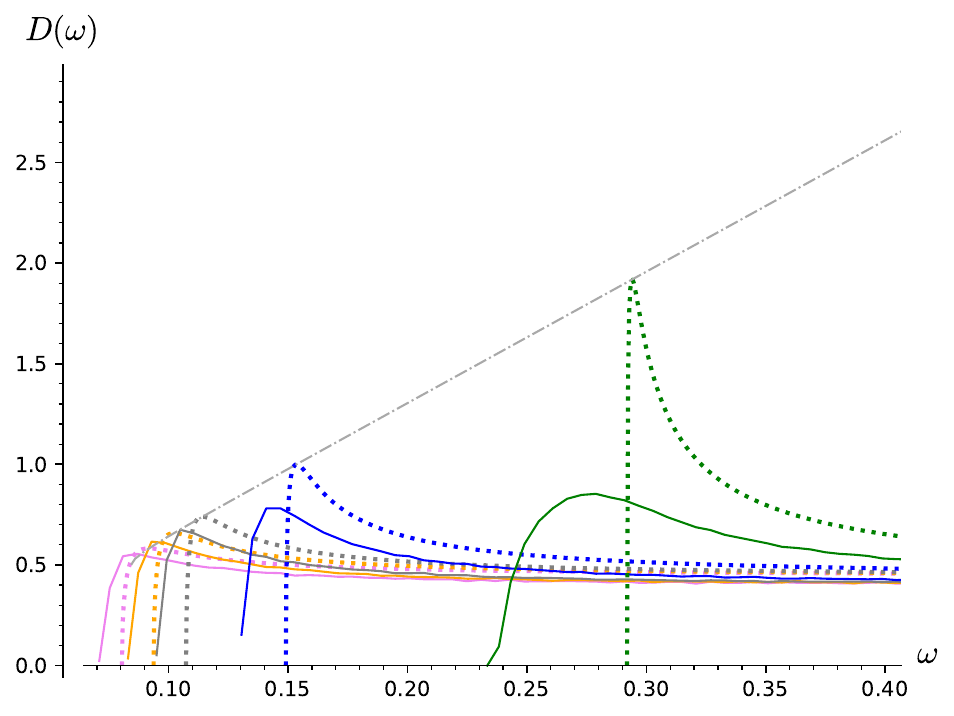,  width=5.00cm}
\caption{
Plots of simulations of the DOS in $d=3$ and $t=2.1$ for $\omega \le 0.4$,
for $R=0.05$ (violet), $0.06$ (orange), $0.07$(grey), $0.1$ (blue) and $0.2$ (green), 
with $500$ random graphs with $1000$ nodes.
	The grey dashdot line is the line of the top of the peaks, see Eq. (\ref{omp}).
The dashed lines represent the infinite dimensional solution.
	}
\label{figt21}
\end{center}
\end{figure*}

	For $t-2$ small and positive, there is a peak in the DOS before the plateau,
when $R > \sqrt{1 + \sqrt{2}}\, R_0$;
in Fig. \ref{figt21} we plot $D^{(R^2, t)}_d$ for $t=2.1$, 
$d=3$ and $R=0.05, 0.06, 0.07, 0.1$ and $0.2$.
As $R$ increases, the difference between the peak in $D^{(R^2, t)}_d$ and the
one in the infinite dimensional solution increases.

\begin{figure*}[hbt!]
\begin{center}
\epsfig{file=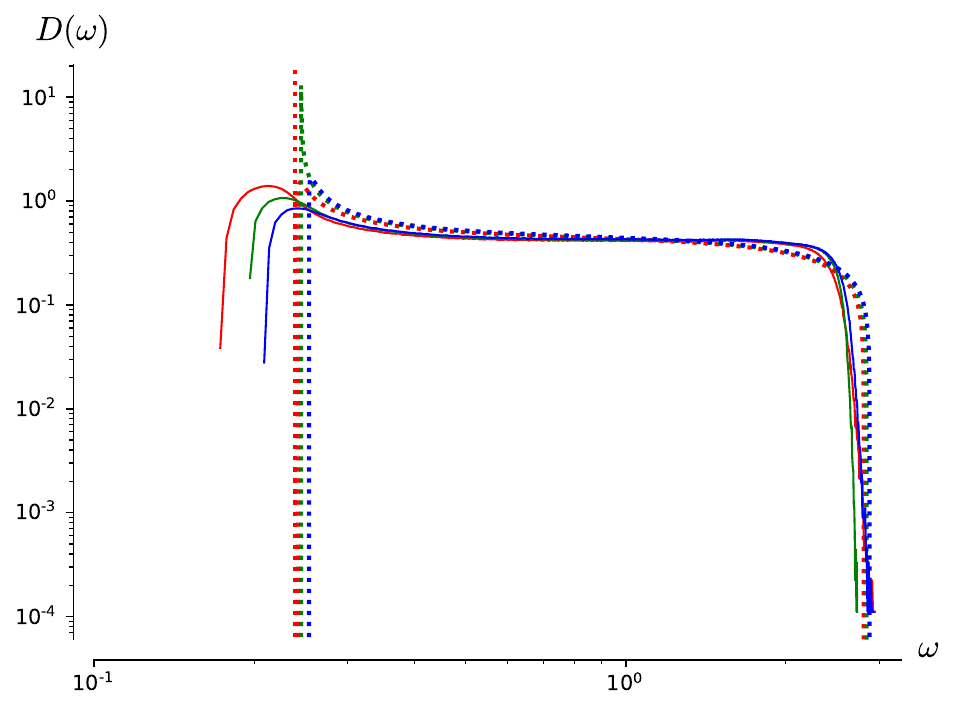,  width=7.00cm}
	\caption{Plots of simulations of the DOS in $d=3$ and $R = 0.1732$ 
for $t=1.9$ (red), $2$ (green) and $2.1$ (blue), with $500$ random graphs
with $1000$ nodes.
In the case $t=1.9$ a random graph is the sum of a regular random graph of degree $5$
and of an Erd\"os-Renyi graph of degree $0.7$.
In the case $t=2$ regular random graphs of degree $6$ are used.
In the case $t=2.1$ a random graph is the sum of a regular random graph of degree $6$
and of an Erd\"os-Renyi graph of degree $0.3$.
The dashed lines represent the infinite dimensional solution.
	}
\label{pe2}
\end{center}
\end{figure*}

In Fig. \ref{pe2} we plot the DOS in the case $d=3$, $R = 0.1732$, corresponding 
to $e = -\frac{R^2}{d} \approx -0.01$
in the notation of \cite{degiuli}. Comparing with  Fig. $13$ in that paper,
we get a peak before the plateau, but no gap for $\delta Z = -0.3$.

\newpage
\section{Conclusion}
We studied a random Laplacian block matrix modeling the Hessian of an amorphous
solid with connectivity close to the isostatic point and applied stress.
In an infinite dimensional limit the spectral distribution is a translated
Marchenko-Pastur distribution. With positive applied stress there is a stability
region; for negative applied stress the density of states has a delta-function peak 
in the hypostatic case, a peak before the plateau in the DOS in the hyperstatic case
for large enough applied stress amplitude.

We made some numerical simulations in low dimensions.
Jammed elastic spheres are not stable under compression in this model:
in the case of positive applied stress
there are a few negative eigenvalues of the Hessian matrix,
which become less frequent increasing the dimension.
In the case of negative applied stress the peak before the plateau is
broader and lower than in the infinite dimensional limit; the difference with
the infinite dimensional limit increases with the applied stress.
It would be interesting to investigate this model with the cavity method
as done in \cite{benet} in absence of applied stress.

\section{Acknowledgments}
I thank G. Cicuta for discussions.

 \end{document}